\documentclass[journal]{IEEEtran}
\pdfoutput=1
\usepackage[noadjust]{cite}
\usepackage{graphicx}
\usepackage{epstopdf}
\usepackage{color}
\usepackage{placeins}
\usepackage{float}
\usepackage{tabularx,colortbl}
\usepackage{cite}

\usepackage{textcomp}
\usepackage{amsmath,amssymb}
\usepackage{xcolor}

\usepackage{graphicx}
\usepackage{epstopdf}
\usepackage{epstopdf}
\usepackage{cite}
\usepackage{amsmath,amssymb}
\newcommand\numberthis{\addtocounter{equation}{1}\tag{\theequation}}
\usepackage{xcolor}
\usepackage{textcomp}

\usepackage{flushend}

\begin{document}
\bstctlcite{IEEEexample:BSTcontrol} % adds et. al. when more than 3 authors (also commands in .bib)

\title{Millimeter-Wave Angle Estimation of Multiple Targets Using Space-Time Modulation and Interferometric Antenna Arrays}
\author{Stavros Vakalis,~\IEEEmembership{Student Member,~IEEE} and Jeffrey A. Nanzer,~\IEEEmembership{Senior Member,~IEEE}%

\thanks{Manuscript received 2020.}
\thanks{This material is based upon work supported by the National Science Foundation under grant number 1751655. \textit{(Corresponding author: Jeffrey A. Nanzer)}}
\thanks{The authors are with the Department of Electrical and Computer Engineering, Michigan State University, East Lansing, MI 48824 USA (email: \{vakaliss, nanzer\}@msu.edu).}
} % stops a 

\maketitle

\begin{abstract}
A new method of angle estimation of multiple targets using a distributed interferometric antenna array and wideband space-time modulation is presented in this work. Interferometric array measurements of angle of arrival are generally ambiguous in the presence of one or more targets. We propose a new method of mitigating ambiguities in interferometric measurements by multiplying the angle pseudo-spectra from multiple antenna baselines, resulting in detections at only the angles of the targets. Using a single linear frequency modulated (LFM) transmitter and a receive interferometric array with $N$ elements we show a simple and computationally efficient technique to estimate the angle of up to $\mathcal{O}(N^2)$ targets. We describe the theory behind the technique, present detailed simulations, and an experimental verification using a three-element millimeter-wave measurement system.
\end{abstract}

\begin{IEEEkeywords}
Angle estimation, correlator arrays, interferometric antenna arrays, interferometry, radar, space-time modulation, target detection \end{IEEEkeywords}

\IEEEpeerreviewmaketitle

\section{Introduction}

\IEEEPARstart{R}{adar systems} performing angle estimation have traditionally been used in applications that require all-weather capabilities for monitoring a spatial environment, such as air traffic control \cite{Skolnik2001, 249135}, automotive radar \cite{6127923}, intruder detection \cite{6494404}, and wireless sensors networks \cite{4357952, 4435061}. Compared to the measurement of relative velocity and range, angle estimation is a more complicated measurement to implement with good accuracy, and traditionally has required some form of spatial scanning or digital receive array computational techniques such as eigenvalue decomposition. Mechanical scanning systems with rotating gimbals represent a simple and cost-effective solution for angle estimation. These systems however tend to be bulky, too slow for many applications, require additional power for mechanical gimbals, and are limited by the angular resolution of the antenna. A way to increase the analog beam-scanning speed is to use electronically scanning phased arrays \cite{6547987, 6898043} which can achieve improved gain, spatial selectivity, and significantly faster scanning speeds than mechanically actuated systems. Nevertheless, phased antenna arrays require a large number of active components and a multi-element filled aperture which results in high cost \cite{7544459}. The speed for electronically beam-scanning can furthermore be restricting because all elements in an electrical scanning phased array are capturing information associated with one angle at a time. 

Estimating target angle without beamscanning can significantly increase the capabilities of remote sensing systems.
Subspace techniques such as MUSIC \cite{1143830} and ESPRIT \cite{32276} have been extensively used for angle estimation without beamscanning, and can achieve resolution smaller than the diffraction limit of the aperture. However, these approaches leverage \textit{a priori} information about the number of sources, which is not always available, and using $N$ antenna elements they cannot in general estimate the angle of more than $N-1$ sources. 
Multiple-input multiple-output (MIMO) radar 
%can achieve improved target identifiability of an $N$-element receive array by 
using an $M$-element transmit array with diverse transmit waveforms and antenna spacings \cite{4350230} can resolve the angle of number of targets on the order of up to $\mathcal{O}(M N)$ \cite{4358016}. This improvement comes with the additional hardware of a transmit array, which is not only costly but also power hungry, and increases the total system complexity. Co-prime and nested array processing can combine the information from two antenna arrays with $M$ and $N$ elements respectively in order to achieve target identiability on the order of $\mathcal{O}(M N)$ in a completely passive setting \cite{5456168, 5609222}. This removes the need for expensive active hardware but can require computationally expensive signal processing and additional information such as higher order statistics. 

%In this article, 
We present a novel technique for estimating the angle of multiple targets using a single transmitter and an $N$-element interferometric receive array. Using a linear frequency modulated (LFM) transmit waveform, the pair-wise cross-correlation of the signals received by the array result in a space-time modulated radiation pattern that imparts a deterministic angular signature as a function of angle. By combining multiple baseline responses, the angle of up to 
%This new method is capable of accurately estimating the angles of up to 
%we will use an $N$-element sparse interferometric array and a single transmit element in order to perform angle estimation of up to 
$\mathcal{O}(N^2)$ targets
can be accurately estimated.
The technique uses interferometric processing that can achieve fast and accurate measurements with a simple hardware architecture. Unlike many direction finding phase interferometry techniques, most of which are passive and only estimate the angles of emitting targets \cite{7277022, intAOA, 8789717}, our approach actively transmits a signal, thus it is possible to estimate the angle of any reflecting targets, not just emitting targets. However, because interferometric techniques can be ambiguous, we propose an approach to filter out the unwanted ambiguities by multiplying the responses from multiple baselines. Furthermore, unlike beam-scanning techniques that focus all the radiation at an angular direction at every instance, interferometric arrays capture information from the entire angular space simultaneously, which therefore significantly shortens the data acquisition time. By using only a single transmitter, our approach requires significantly less hardware than MIMO radar, and our signal processing approach based on vector multiplications and fast Fourier transforms is fast and efficient. Previously we described the general operational concept and presented simulated results \cite{aps2020}. In the following, we describe in detail the theory behind interferometric angle estimation and the approach for mitigating unwanted ambiguous target responses. We present detailed simulations of angle estimation, and assess the angle estimation root-mean square error (RMSE) and probability of false alarm ($P_{FA}$). Finally, we present a proof-of-concept demonstration using a three-element millimeter-wave experimental system.

\begin{figure}[t]
	\begin{center}
		\noindent
		\includegraphics[width= 2.3in]{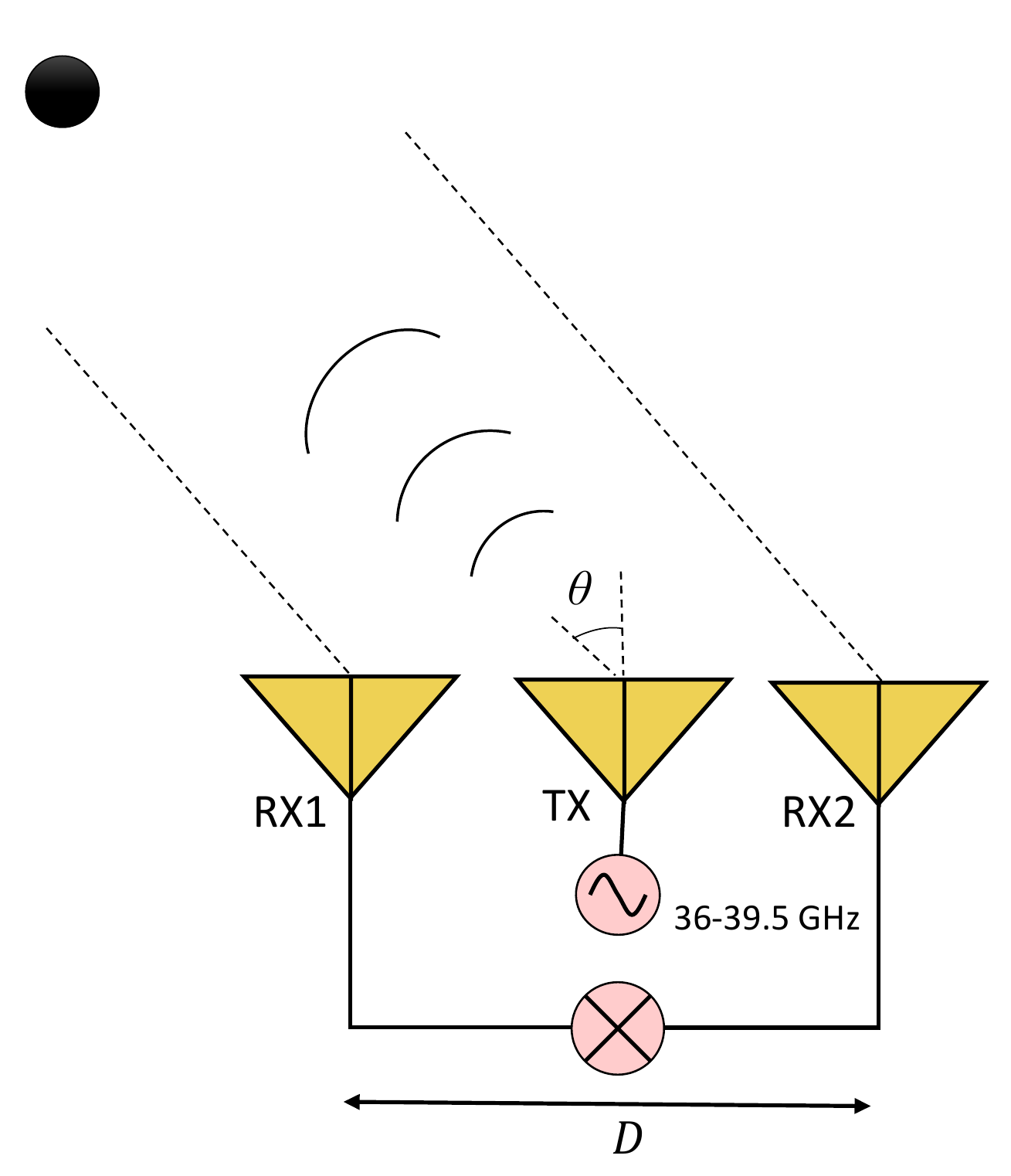}
		\caption{Active distributed array observing a target at angle $\theta$. The transmitter generates an LFM waveform over the 36-39.5 GHz band, which is received by the two receiver antennas separated by the baseline $D$. }
		\label{fig:fig1}
	\end{center}
\end{figure}

\section{Interferometric Angle Estimation Using Space-Time Modulated Array Patterns}
%\subsection{Theory}
Simple and accurate angle estimation can be accomplished using a distributed array and wideband spatial frequency modulation \cite{8493305}. A single transmitter emitting an LFM signal and two receive antennas separated by a baseline $D$ can be seen in Fig. \ref{fig:fig1}. The transmit signal is reflected from a target at angle $\theta$ and the two received signals are cross-correlated which is realized by multiplication and low-pass filtering. %This process can be implemented with analog hardware or digitally
The transmitted LFM signal $s(t)$ can be written as
%A pair of antennas separated by a baseline D, as shown in Fig. \ref{fig:fig1}, forms a correlation interferometer. Assuming that the transmitted signal is an LFM chirp of the form
\begin{equation}\label{eq.tr}
s(t) = \cos \left[2 \pi \left( f_0 t + \tfrac{K}{2} t^2 \right) \right]
 \end{equation}
where $f_0$ is the carrier frequency, and $K$ (Hz/s) is the chirp rate. The in-phase received signals on the two antennas separated by a baseline $D$ after reflecting off of a point source residing at angle $\theta$ can be given by
 %and that both antennas are receiving signals from the same radiating source, then the normalized output at the receivers is given by
 \begin{align}
v_1(t) &= \cos \left[2 \pi \left( f_0 t + \tfrac{K}{2} t^2 \right) \right])\label{eq.V1}+n_1(t)\\
v_2(t) &= \cos \left\{2 \pi \left[ f_0 \left(t-\tau_g \right) + \tfrac{K}{2} \left(t-\tau_g \right)^2 \right] \right\} +n_2(t)\label{eq.V2}
\end{align}
where $ \tau_g = \frac{D}{c}\sin \theta$ represents the geometrical time delay that the wavefront experiences between reaching the two antenna elements and $n_i(t)$ is the noise on the $i$th receiver. The in-phase part of the cross-correlation of the two received signals can be written as
 \begin{align*}
r^I(\tau_g) &= \langle v_1(t) v_2(t) \rangle \\
		&= \Big\langle \cos \left[2 \pi \left( f_0 t +  \tfrac{K}{2} t^2 \right) \right] \times \\
		&\phantom{{}=1} \ \cos \Big{\{}2 \pi \left[ f_0 \left(t-\tau_g \right) +  \tfrac{K}{2} \left(t-\tau_g \right)^2 \right] \Big{\}}  \Big\rangle \\
		&= \cos \left[ 2 \pi \left( f_0 + Kt - \tfrac{K}{2} \tau_g \right)\tau_g \right] \numberthis \label{eq.ri}
\end{align*}
where the angled brackets $\langle \cdot \rangle$ indicate time-averaging, which can be accomplished with low-pass filtering. The noise terms do not show up in \eqref{eq.ri} because they are uncorrelated with each other and with the received signal, and thus tend to zero with sufficiently long integration time (or, equivalently, sufficiently low cutoff frequency in the low pass filter). Low pass filtering also removes the higher frequency carrier components. 
\begin{figure}[t!]
	\begin{center}
		\noindent
		\includegraphics[width= 3.1in]{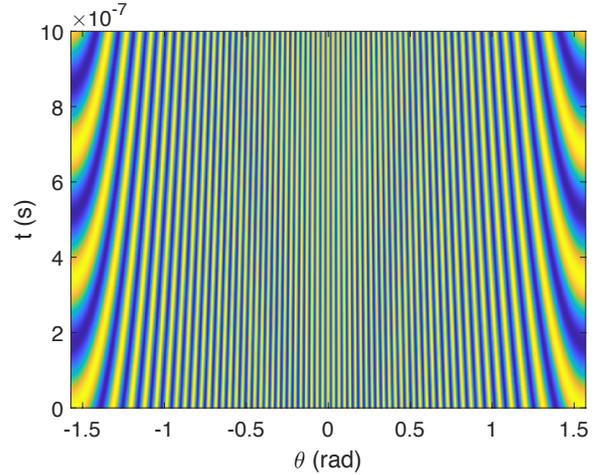}
		\caption{Spatio-temporal pattern of a correlation interferometer with $D$=0.25 m and frequency ranging from 36 to 39.5 GHz. As time increases more grating lobes appear in the spatial domain, introducing a space-time modulation that yields a unique signature as a function of angle.}
		\label{fig:fig2}
	\end{center}
\end{figure}

\begin{figure}[t]
	\begin{center}
		\noindent
		\includegraphics[width= 2.6in]{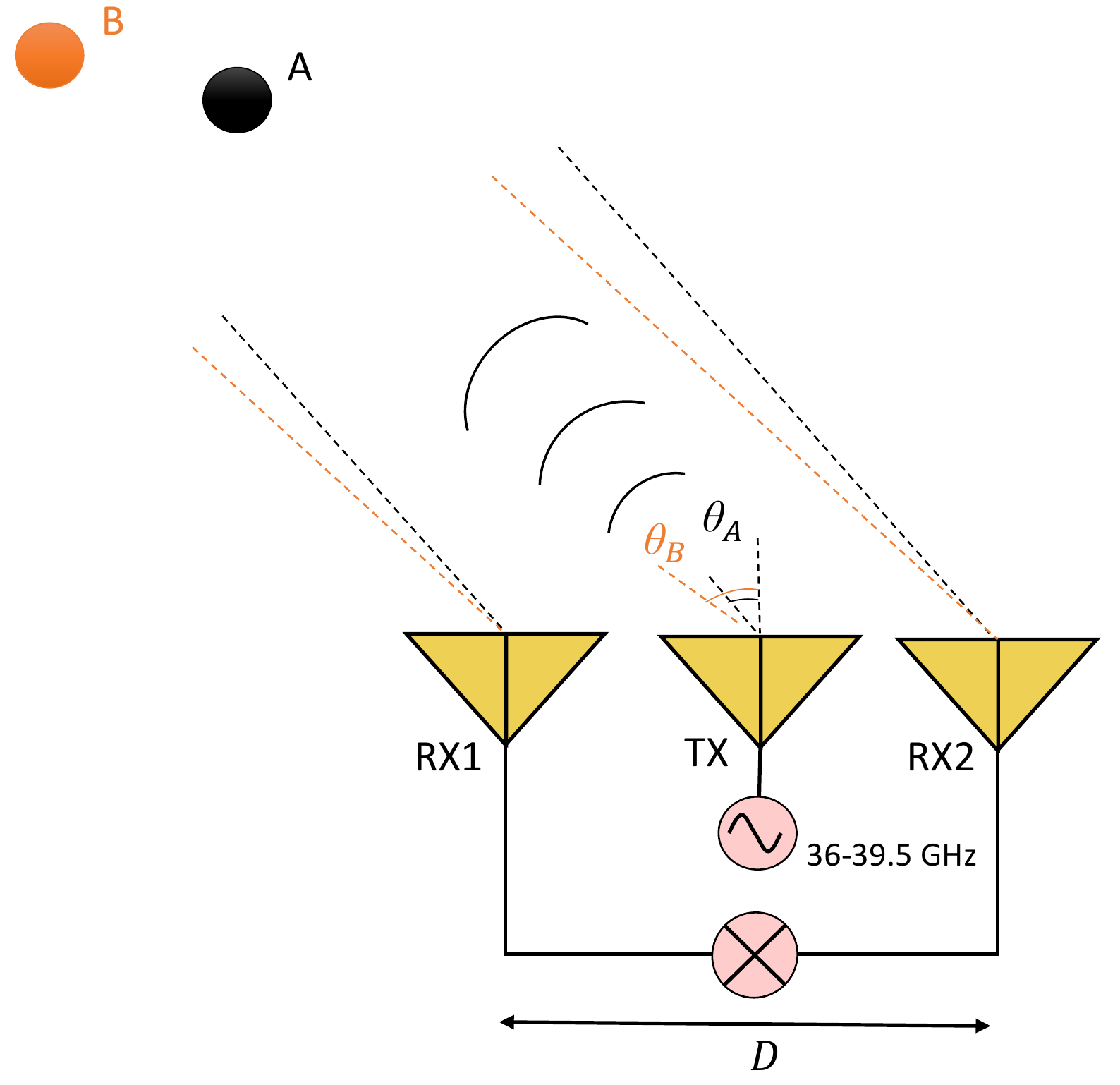}
		\caption{Active distributed array observing two targets $A$ and $B$ at angles $\theta_A$ and $\theta_B$.}
		\label{fig:fig3}
	\end{center}
\end{figure}

\begin{figure*}[t!]
	\begin{center}
	\noindent
		\includegraphics[width=7 in]{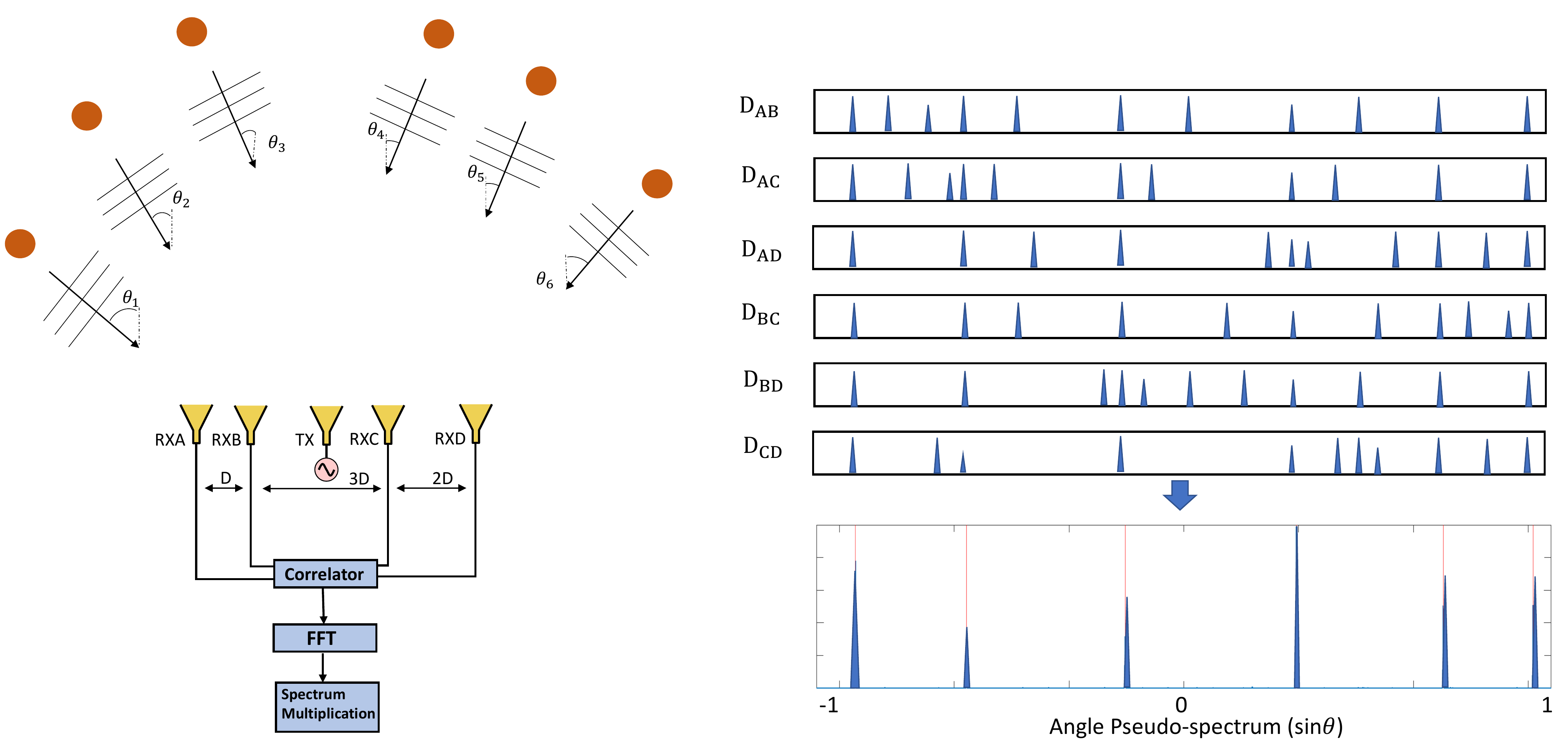}
  			\caption{(Left) $N=$ 4-element receive array with a single transmitter observing the reflections from $\tfrac{1}{2}N(N-1)= 6$ targets. The received signals are cross-correlated pairwise and the cross-correlation outputs are Fourier transformed to convert them to angle pseudo-spectra, which are multiplied to keep only the useful angle information. (Right) Visualization of the 6 individual baselines which contain both the useful and unwanted frequency information. The bottom plot shows the multiplicative correlation response, which retains only the signal responses at the true target angles.}
		\label{array_conf}
	\end{center}
\end{figure*}

Combining the in-phase and quadrature outputs of the cross-correlation produces the complex signal response
\begin{equation}
r(\tau_g) = e^{ j  2 \pi \left( f_0 + Kt - \tfrac{K}{2} \tau_g \right) \tau_g  }.
\end{equation}
The instantaneous frequency $f_i$ of the receiver circuit output can be found from the derivative of the phase through
 \begin{equation}\label{eq.psi}
 f_i=\frac{1}{2\pi} \frac{d\phi}{dt}= K \tau_g = K\frac{D}{c}\sin \theta.
\end{equation}
The target angle is then obtained through
 \begin{equation}\label{eq.psi2}
 \theta = \sin^{-1}\left(\frac{f_i c}{KD}\right).
\end{equation}
The instantaneous frequency is proportional to the chirp rate $K$, the baseline $D$, and the sine of the angle $\theta$ at which the reflecting target resides. For angles $\theta \in [-\frac{\pi}{2}, \frac{\pi}{2}]$, each angular point has a unique frequency response as a function of time. Fig. \ref{fig:fig2} shows the spatio-temporal transmit pattern of a correlation interferometer as a function of time and one angular dimension. It can be seen that more grating lobes manifest as time increases. This is because the instantaneous frequency of the LFM signal increases linearly, and as a result the electrical baseline for two fixed antenna elements follows the same linear response. A two-element correlation interferometer with linearly increasing carrier frequency introduces a space-time modulation the manifests a unique sinusoidal response per angle. This leads to a simple and accurate angle estimation approach by simply estimating the frequency of the received signal, which uniquely matches a specific angle.

The drawback of this approach is that it cannot unambiguously estimate the angle of more than one target. Consider the configuration of a two-element correlation interferometer receiver and a single transmitter observing the reflections from two targets $A$ and $B$ as shown in Fig. \ref{fig:fig3}. Given the transmit LFM signal, the normalized signal responses in the two receive elements can be found as
%A pair of antennas separated by a baseline D, as shown in Fig. \ref{fig:fig1}, forms a correlation interferometer. Assuming that the transmitted signal is an LFM chirp of the form
%\begin{equation}\label{eq.tr}
%s(t) = \cos \left[2 \pi \left( f_0 t + \tfrac{K}{2} t^2 \right) \right]
% \end{equation}
%where $f_0$ is the carrier frequency, and $K$ (Hz/s) is the chirp rate. Assuming the
 %and that both antennas are receiving signals from the same radiating source, then the normalized output at the receivers is given by
 \begin{align*}
v_1(t) &= \cos \left[2 \pi \left( f_0\left(t-\tau_{A1} \right) + \tfrac{K}{2} \left(t-\tau_{A1} \right)^2 \right) \right] +\\   &\cos \left\{2 \pi \left[ f_0 \left(t-\tau_{B1} \right) + \tfrac{K}{2} \left(t-\tau_{B1}  \right)^2 \right] \right\} + n_1(t) \numberthis \label{eq.V11}
\end{align*}
 \begin{align*}
v_2(t) &= \cos \left\{2 \pi \left[ f_0 \left(t-\tau_{A2} \right) + \tfrac{K}{2} \left(t-\tau_{A2} \right)^2 \right] \right\} + \\ 
&\cos \left\{2 \pi \left[ f_0 \left(t-\tau_{B2} \right) + \tfrac{K}{2} \left(t-\tau_{B2} \right)^2 \right] \right\} + n_2 (t) \numberthis \label{eq.V21}
\end{align*}
where the terms $\tau_{Ai}$,  $\tau_{Bi}$ represent the total round trip time delays between the signal transmission and the reflection from target A and B arriving at the $i$th antenna element.  The output of the cross-correlation of the two in-phase signals is then
 \begin{align*}
r^I &= \langle v_1(t) v_2(t) \rangle \\
		&= \Big\langle \Big( \cos \left[2 \pi \left( f_0\left(t-\tau_{A1} \right) + \tfrac{K}{2} \left(t-\tau_{A1} \right)^2 \right) \right] +\\   &\cos \left\{2 \pi \left[ f_0 \left(t-\tau_{B1} \right) + \tfrac{K}{2} \left(t-\tau_{B1}  \right)^2 \right] \right\} \Big) \times \\
		&\phantom{{}=1} \ \Big( \cos \left\{2 \pi \left[ f_0 \left(t-\tau_{A2} \right) + \tfrac{K}{2} \left(t-\tau_{A2} \right)^2 \right] \right\} + \\ 
		&\cos \left\{2 \pi \left[ f_0 \left(t-\tau_{B2} \right) + \tfrac{K}{2} \left(t-\tau_{B2} \right)^2 \right] \right\} \Big) \Big\rangle. \numberthis \label{eq.ri1}
\end{align*}
Combining the in-phase and quadrature correlator outputs results in the complex signal response
\begin{align*}
r(\tau_g) &= e^{ j  2 \pi \left( f_0 (\tau_{A2} - \tau_{A1}) + Kt(\tau_{A2} - \tau_{A1}) - \tfrac{K}{2} (\tau_{A2}^2 - \tau_{A1}^2) \right)} +\\ 
& e^{ j  2 \pi \left( f_0 (\tau_{B2} - \tau_{B1}) + Kt(\tau_{B2} - \tau_{B1}) - \tfrac{K}{2} (\tau_{B2}^2 - \tau_{B1}^2) \right)} + \\ & e^{ j  2 \pi \left( f_0 (\tau_{B2} - \tau_{A1}) + Kt(\tau_{B2} - \tau_{A1}) - \tfrac{K}{2} (\tau_{B2}^2 - \tau_{A1}^2) \right)} + \\ & e^{ j  2 \pi \left( f_0 (\tau_{A2} - \tau_{B1}) + Kt(\tau_{A2} - \tau_{B1}) - \tfrac{K}{2} (\tau_{A2}^2 - \tau_{B1}^2) \right)}.  \numberthis \label{eq.rout}
\end{align*}

The instantaneous frequencies of the four terms in \eqref{eq.rout} can be found from the derivative of the phase to be $K(\tau_{A2} - \tau_{A1})$, $K(\tau_{B2} - \tau_{B1})$, $K(\tau_{B2} - \tau_{A1})$, and $K(\tau_{A2} - \tau_{B1})$ respectively. The first two terms $K(\tau_{A2} - \tau_{A1})$ and $K(\tau_{B2} - \tau_{B1})$ correspond to the geometric time delay terms $K\frac{D}{c}\sin \theta_A$ and $K\frac{D}{c}\sin \theta_B$ and they represent the useful information that we need to capture in order to estimate the angle of the two targets. The last two terms $K(\tau_{B2} - \tau_{A1})$ and $K(\tau_{A2} - \tau_{B1})$ cannot be matched to any target angle and they cannot be differentiated from the actual frequency responses that represent target angles. These terms come from the difference in round trip times of the signal reflecting from target $B$ to reach antenna element 2 and the response from target $A$ to receive antenna element 1, and vice versa. In general when capturing the reflections from $N$ targets, up to $N^2$ different frequency responses may manifest in the correlator output. Thus for $N$ targets on different angles we will have $N$ desired frequency responses and up to $N(N-1)$ undesired frequency responses. The next section describes a multi-baseline approach to mitigate these false angle responses.
%In the next section we will discuss a way to keep only the useful angle information.

\section{Disambiguation Using Multiple Interferometric Baselines}
An instantaneous frequency response of a target at angle $\theta$, given by \eqref{eq.psi}, can have only values such that $f_i$~$\in$~$[-K\frac{D}{c}, K\frac{D}{c}]$ for $-1 \leq \sin \theta \leq  1$. So for two targets $A$ and $B$ and for a two element interferometer if $(\tau_{B2} - \tau_{A1})$ and $(\tau_{A2} - \tau_{B1})$ are both larger than $D/c$ then they could simply be ignored as a potential target response because they do not correspond to a real angle. This approach will potentially filter out some unwanted information, especially for small baselines $D$. However, false target angles will generally not always manifest at angles outside of the real range, especially for a large number of targets $N$ where the number of unwanted cross-product frequency responses can be up to $N(N-1)$. 

Our solution starts with the use of an interferometric array instead of a single correlation interferometer. An antenna array with $N$ elements that uses interferometric processing can have up to $\tfrac{1}{2}N(N-1)$ correlation interferometers with unique baseline spacings. Actual target responses will manifest at the same locations of the angle pseudo-spectrum in all different baselines of the array, however the unwanted cross-product terms will not manifest on all antenna baselines at the same locations of the angular spectrum. Thus, the desired true angle information will appear consistently in all different baselines, however this is not the case for the cross-products terms where the false angle signals will generally be low in amplitude. We thus multiply the responses of the multiple baselines, resulting in large signal power for the true target signals that are consistent in angle across all baselines, and reducing the signal power of the false angle targets, at which angles most baselines have a low signal amplitude. 

The algorithm can be visualized in Fig. \ref{array_conf}. The receive antenna elements capture the reflections of the transmitted signals from the targets in front of the array. The responses of each element are captured, multiplied pair-wise, and low pass filtered to produce the $\tfrac{1}{2}N(N-1)$ correlation interferometer responses. The output of each correlation interferometer is then Fourier transformed and converted to the angle spectrum using \eqref{eq.psi2}. By multiplying the angle spectra, only the common responses are retained from the $\tfrac{1}{2}N(N-1)$ responses, since for most baselines the false angle responses are of low amplitude.
%That is because even if a undesired response can appear in multiple baseline spectrum at the same angle, if at least one baseline do not have this undesired angular response, its amplitude at this location will be close to zero. 
The process thus works like a filter in the angular pseudo-spectrum. In the next section we demonstrate through simulation the ability to accurately estimate multiple target angles for arrays with $N$ elements when observing number of targets even larger than $\tfrac{1}{2}N(N-1)$ under various signal-to-noise ratio (SNR) scenarios.

\section{Simulated Angle Reconstructions of $\mathcal{O}(N^2)$ Targets}
\begin{figure}[t!]
	\begin{center}
		\noindent
		\includegraphics[width= 3.5in]{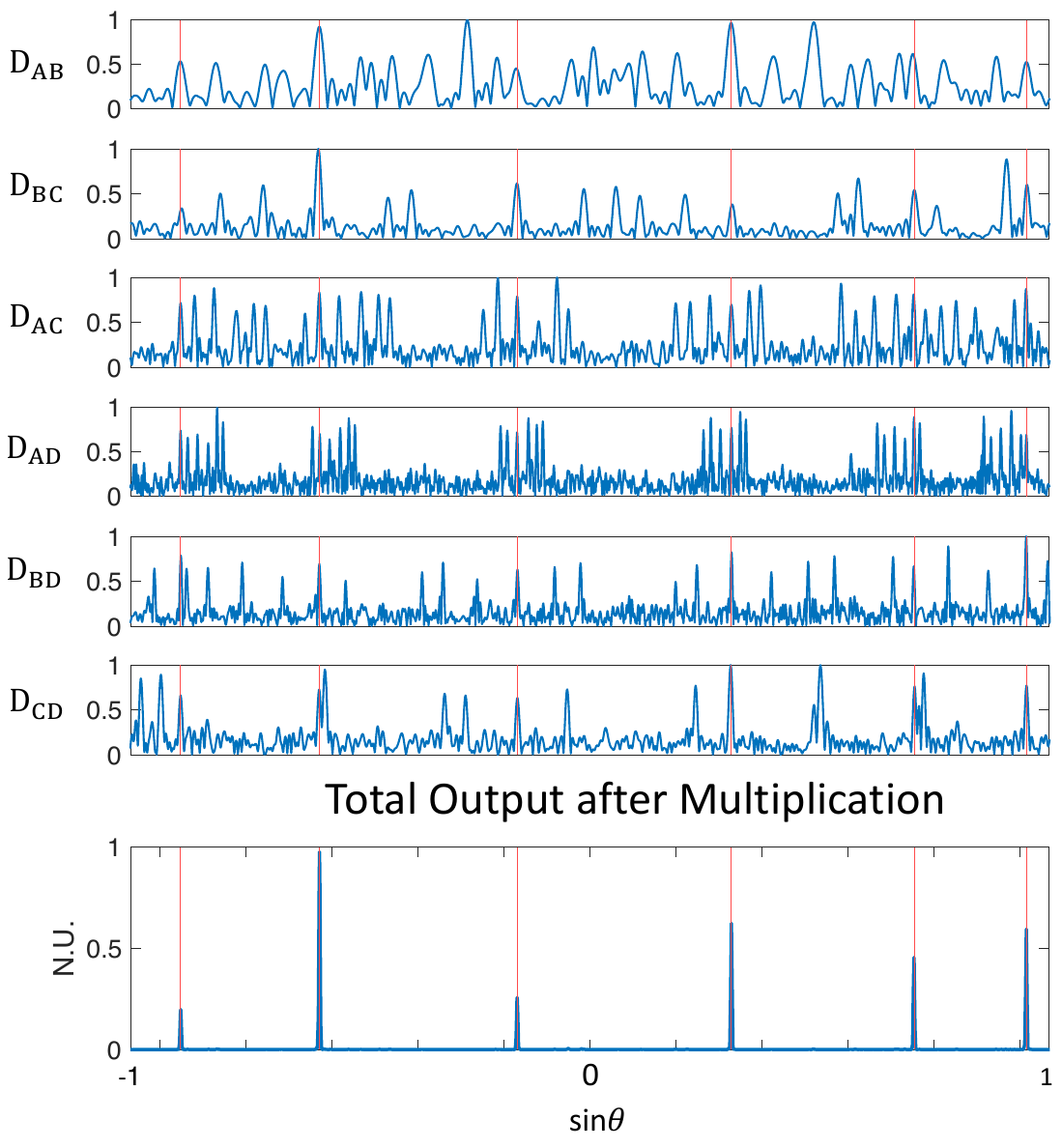}
		\caption{Simulated results for a 4-element receive interferometric array with elements $\{A,B,C,D\}$ in spacings $\{2D, 3D, 5D \}$observing the responses from $\tfrac{1}{2}N(N-1)= 6$ targets, with SNR = 10dB. Although each baseline has unwanted cross-product terms, in the bottom it can be seen that only the responses from the six targets remain.}
		\label{fig:sim}
	\end{center}
\end{figure}

\begin{figure}[t!]
	\begin{center}
		\noindent
		\includegraphics[width= 3.5in]{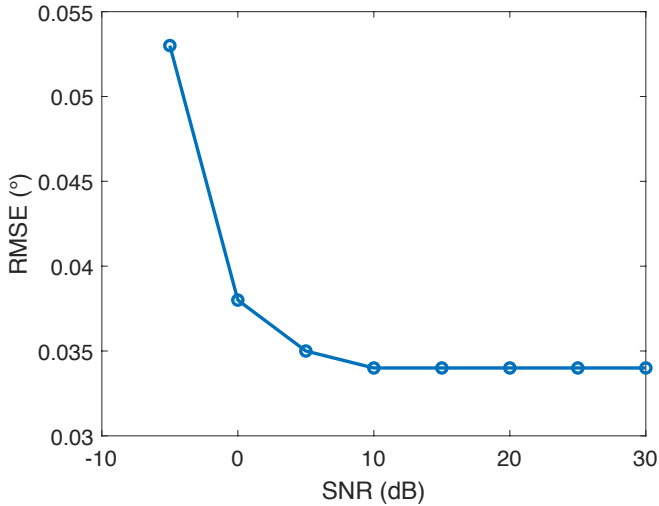}
		\caption{Root-mean-square error (RMSE) plot for a 4-element array observing the responses from $\tfrac{1}{2}N(N-1)= 6$ targets. Due to the correlation process, this technique works well even at negative SNR values.}
		\label{fig:rmse500}
	\end{center}
\end{figure}

\begin{figure}[t!]
	\begin{center}
		\noindent
		\includegraphics[width= 3.5in]{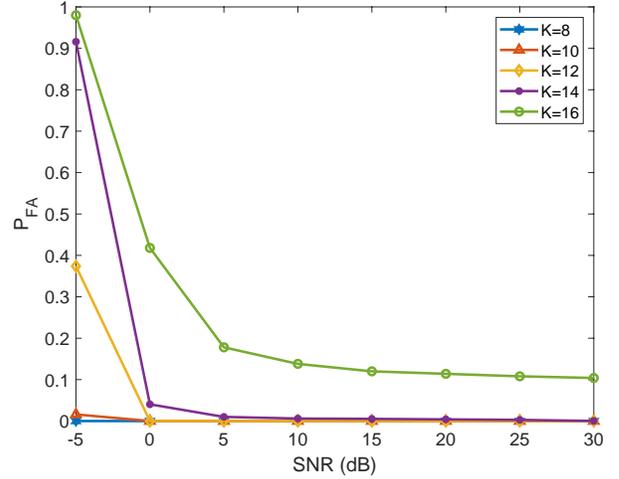}
		\caption{Probability of false alarm ($P_{FA}$) for a 5-element array observing the reflections from $K$ targets. The technique shows promising results even for $K\geq \tfrac{1}{2}N (N-1)$.}
		\label{fig:pfa}
	\end{center}
\end{figure}
Simulations were run in MATLAB for an antenna array  with $N=4$ elements $\{A,B,C,D\}$ with spacings $\{2D, 3D, 5D \}$ where $D=1$ m. An SNR of 10 dB was implemented. The bandwidth of the LFM was 4 GHz, and the pulse repetition interval was  1 $\mu$s, resulting in a chirp rate $K=4 \cdot 10^{15}$ Hz/s. The response from $\tfrac{1}{2}N(N-1)= 6$ targets at angles $-63^\circ$, $-36^\circ$, $-9^\circ$, $18^\circ$, $45^\circ$, and $72^\circ$ with equal reflectivities can be seen in Fig. \ref{fig:sim}. The individual correlator responses each show signals at the target angles (represented by red lines) and a number of false responses at other angles. The resulting multiplicative correlator response retains only the true target angle responses. We note that while the reflectivities of each target were identical, the resultant amplitudes of the signals in the multiplicative correlator response are not; this is due to the interference generated by false target responses in some of the individual correlator responses. Nonetheless, the resultant responses of the true target angles show significantly higher signal stranegth than the noise floor, even at 10 dB SNR, leading to robust detection. Furthermore, we note that typical array-based angle estimation techniques also do not reconstruct target amplitudes \cite{1143830,32276}.

Monte Carlo simulations were run for the same 4-element array observing the angle of 6 targets for 500 iterations and calculating the RMSE for different SNR values. The results are shown in Fig. \ref{fig:rmse500}, which show that the RMSE is constant for SNR values above 10 dB, and is less than 0.04$^\circ$ for 0 dB SNR. Although there is no coherent processing such as matched filtering, the cross-correlation applies some denoising to uncorrelated noise which makes angle estimation possible even for SNR less than 0 dB, with RMSE less than 0.055$^\circ$ for -5 dB SNR. 
%The technique is noise tolerant and it can be seen due to that there are no significant reduction in RMSE for SNR bigger than 0 dB. 

With a large number of false target angles manifesting with increasing baselines and increasing number of targets, it is relevant to study this effect on the multiplicative correlator output. The probability of false alarm ($P_{FA}$) was thus evaluated for a 5-element receive array observing the reflection from $K=8$ to $K=16$ targets under various SNR scenarios. The array had spacings $\{2D, 3D, 5D, 7D\}$ and $D= 5$ m. The $K$ targets were uniformly placed in the $[ -180^\circ, 180^\circ ]$ space and a random offset was applied at each Monte Carlo loop. The results of the simulation can be seen in Fig. \ref{fig:pfa}. It can be seen that for positive SNR values the $P_{FA}$ becomes negligible for number of targets up to $K=14$, which is larger than the number of baselines $\tfrac{1}{2} N (N-1)=10$. The ability to accurately estimate the angles of more targets than baselines in the array is significant, if not intuitive. However, we note that in this approach we are not limited by a system with $\tfrac{1}{2} N (N-1)$ equations, thus it is not unreasonable to be able to solve for more than $\tfrac{1}{2} N (N-1)$ unknowns. Essentially, the approach simply filters out unwanted responses, similar to a microwave filter removing spurious frequency responses, but in a dynamic way. 

\section{Millimeter-Wave Experimental Measurements}
\begin{figure}[t!]
	\begin{center}
		\noindent
		\includegraphics[width=3.5in]{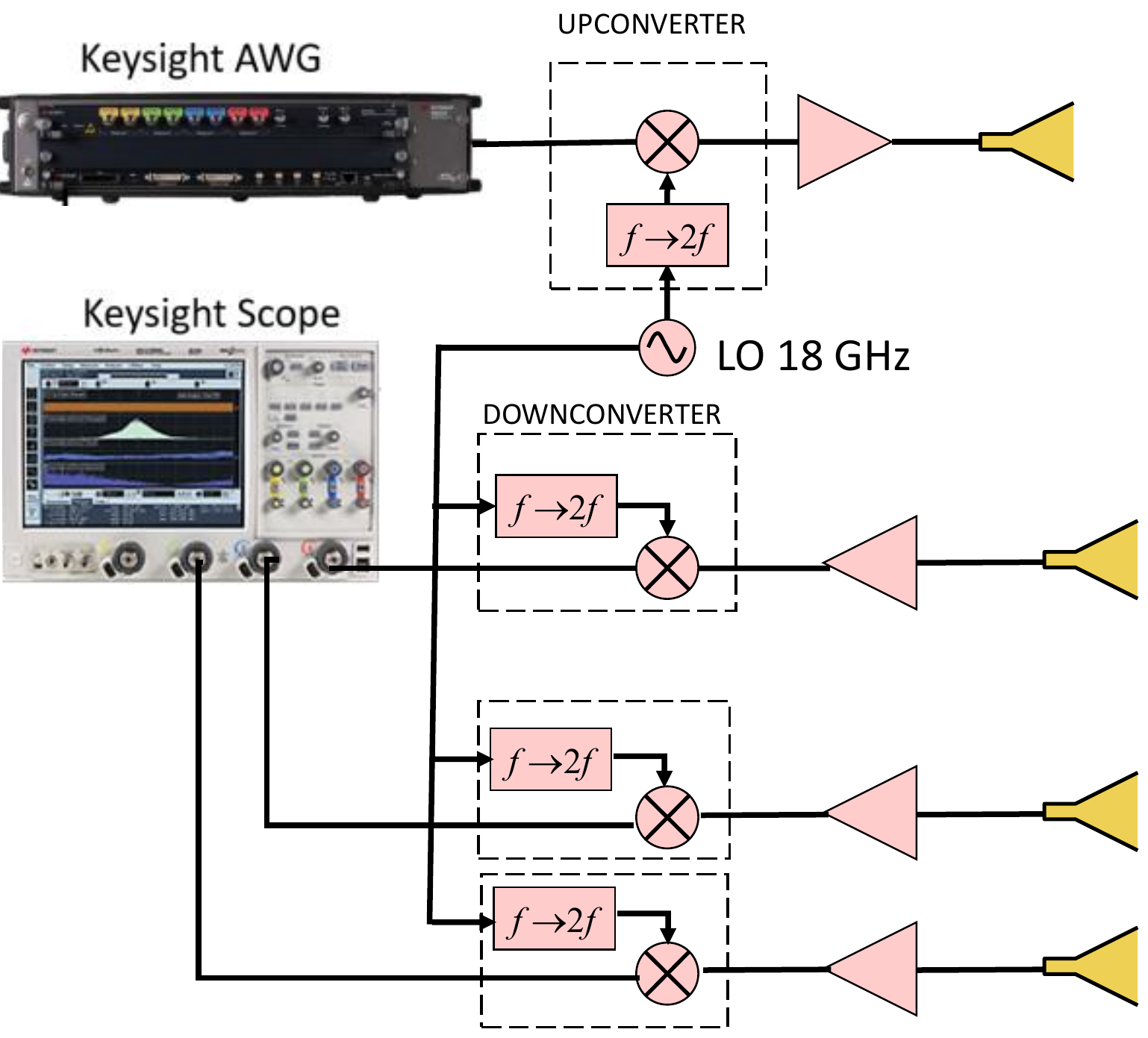}
		\caption{Experimental 36-39.5~GHz configuration with one transmitter and three receivers. The transmit signal was generated at baseband using a Keysight M8190 Arbitrary Waveform Generator (AWG) and the received signals were captured using a Keysight MSOX92004A  mixed-signal oscilloscope.}
		\label{fig:exp_setup0}
	\end{center}
\end{figure}
\begin{figure}[t!]
	\begin{center}
		\noindent
		\includegraphics[width=3.5in]{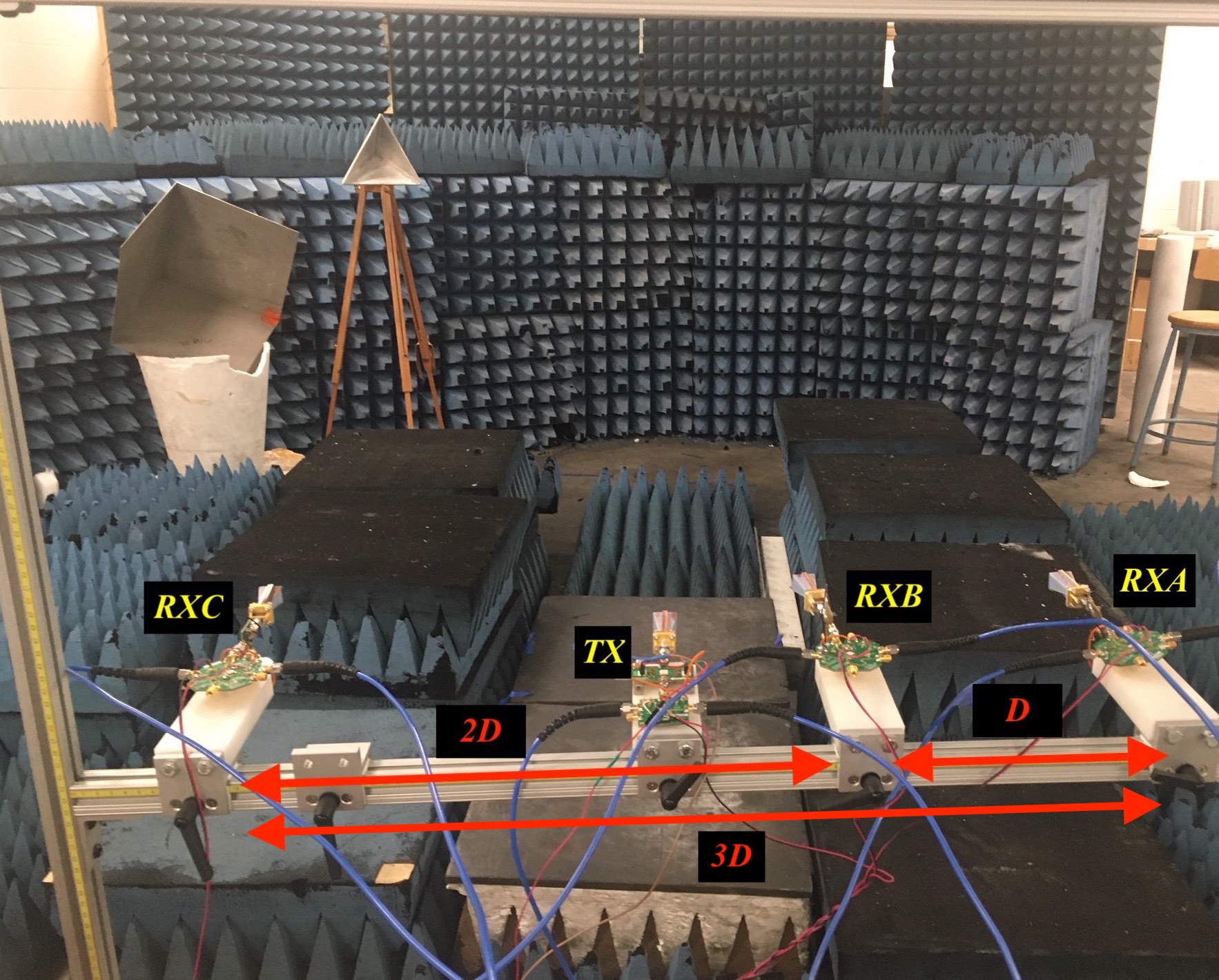}
		\caption{Experimental measurements took place in a semi-anechoic environment with two corner reflectors at $8^\circ$ and $15^\circ$.}
		\label{fig:exp_setup}
	\end{center}
\end{figure}

\begin{figure}[t!]
	\begin{center}
		\noindent
		\includegraphics[width= 3.3in]{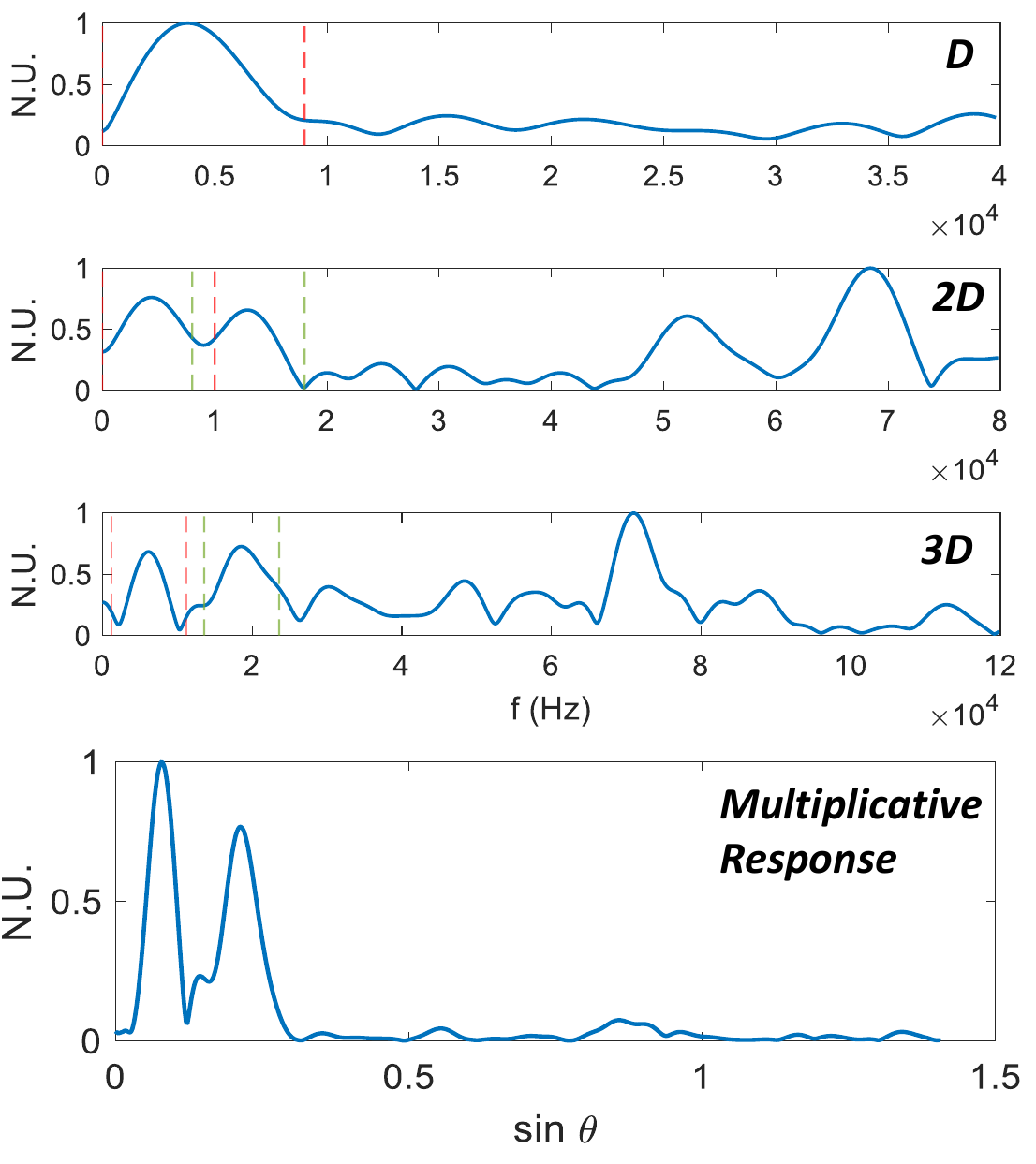}
		\caption{(Top) Experimental frequency responses from three interferometers with baselines $D$, $2D$, and $3D$ observing two targets at angles $8^\circ$ and $15^\circ$. The vertical dotted red and green lines represent the limits of the frequency bins around the target responses. (Bottom) Total system output after multiplying the three individual baseline responses. The unwanted cross-product information has been mitigated significantly.}
		\label{fig:exp_results}
	\end{center}
\end{figure}

%\section{Experimental Measurements}
To verify our approach for angle estimation we built an experimental configuration, a schematic of which can be seen in Fig. \ref{fig:exp_setup0}. We used 15 dBi 3-D printed horn antennas that were fabricated in house. The 3-D printed horns were designed in ANSYS High Frequency Structure Simulator (HFSS) and were printed using VeroWhitePlus material on  Stratus Objet Connex 350 Multi Material 3-D Printing System. For metallization, the structure was sputtered, first with a 60 nm titanium layer for adhesion between the subsequent copper layer and the VeroWhitePlus material. The titanium layer was followed by a 500 nm copper layer. After this step, the sputtered structure was electroplated with copper to achieve a thickness of 5-6 $\mu$m \cite{7836351}.

The transmitted signal was generated at baseband with a Keysight M8190 Arbitrary Waveform Generator (AWG) with a chirp rate of $3.5\times10^{13}$~Hz/s and a pulse duration of 100 $\mu$s. The signal was upconverted with Analog Devices HMC6787ALC5A upconverter at 36-39.5~GHz and then amplified using an Analog Devices HMC7229 24~dB power amplifier. The received signals were amplified by 23 dB gain Analog Devices HMC1040LP3CE low-noise amplifiers, and then downconverted using Analog Devices HMC6789BLC5A downconverters. The up- and
down-converters included a frequency doubler and were driven by a local oscillator (LO) signal of 18~GHz. The received signals were captured using a Keysight MSOX92004A  mixed-signal oscilloscope and were processed offline in MATLAB.
 
Experimental measurements were conducted inside a 7.6~m semi-enclosed antenna range using two corner reflectors (Fig. \ref{fig:exp_setup}) located at $8^\circ$ and $15^\circ$ relative to the array broadside.  Three receive antennas were spaced at baselines of $D $~=~$0.25$~m and $2D$~=~$0.5$~m with the transmit antenna placed in the center of the array. 
The frequency responses of the three correlation interferometers with baselines $D$, $2D$, and $3D$ can be seen in Fig.~\ref{fig:exp_results}. The smaller baseline with length $D$ cannot differentiate the two targets responses, but the baselines with length $2D$ and $3D$ are able to. The red and green dotted lines show the frequency resolution bin around the peaks, which is the inverse of the integration time which in our case was 10~kHz. The peaks on the right side of the baselines $2D$ and $3D$ do not appear in all three of them and do not represent actual target information. In the bottom of Fig. \ref{fig:exp_results} the multiplicative correlation response is shown, where only the true target angles have been retained. Furthermore, due to the multiplicative process, the response of the combination of the signals yields narrower angular signals than any individual correlator response. The unwanted cross-product information has been mitigated significantly. Although $0 \leq \sin \theta \leq 1$, we have not cropped the horizontal axis to show that some filtering takes place from the responses that do not correspond to real angles. No denoising has been applied to the data, and this technique is able to run in real-time, as the signal processing is based on 1-D vector multiplications and Fourier transforms.

\section{Conclusions}
A novel active technique for angle estimation using an LFM transmit waveform and an interferometric array receiver with a space-time modulated radiation pattern has been demonstrated. In contrast to other active angle estimation approaches, no beam-scanning is necessary in this technique. We introduced a method of filtering ambiguous responses by combining the common information from multiple baselines and the simulated and experimental results show good agreement. With our technique an array with $N$ receivers can accurately estimate the angle of more than $\mathcal{O}(N^2) = 0.5N(N-1)$ targets.
\\
%\ifCLASSOPTIONcaptionsoff
 \newpage
%\fi

\bibliographystyle{IEEEtran}

\begin{thebibliography}{10}
\providecommand{\url}[1]{#1}
\csname url@samestyle\endcsname
\providecommand{\newblock}{\relax}
\providecommand{\bibinfo}[2]{#2}
\providecommand{\BIBentrySTDinterwordspacing}{\spaceskip=0pt\relax}
\providecommand{\BIBentryALTinterwordstretchfactor}{4}
\providecommand{\BIBentryALTinterwordspacing}{\spaceskip=\fontdimen2\font plus
\BIBentryALTinterwordstretchfactor\fontdimen3\font minus
  \fontdimen4\font\relax}
\providecommand{\BIBforeignlanguage}[2]{{%
\expandafter\ifx\csname l@#1\endcsname\relax
\typeout{** WARNING: IEEEtran.bst: No hyphenation pattern has been}%
\typeout{** loaded for the language `#1'. Using the pattern for}%
\typeout{** the default language instead.}%
\else
\language=\csname l@#1\endcsname
\fi
#2}}
\providecommand{\BIBdecl}{\relax}
\BIBdecl

\bibitem{Skolnik2001}
M.~I. Skolnik, \emph{Introduction to Radar Systems}.\hskip 1em plus 0.5em minus
  0.4em\relax McGraw-Hill, 2001.

\bibitem{249135}
C.~R. {Rao}, L.~{Zhang}, and L.~C. {Zhao}, ``Multiple target angle tracking
  using sensor array outputs,'' \emph{IEEE Trans. Aerosp. Electron. Syst.},
  vol.~29, no.~1, pp. 268--271, 1993.

\bibitem{6127923}
J.~{Hasch} \emph{et~al.}, ``Millimeter-wave technology for automotive radar
  sensors in the 77 ghz frequency band,'' \emph{IEEE Trans. Microw. Theory
  Tech.}, vol.~60, no.~3, pp. 845--860, 2012.

\bibitem{6494404}
D.~{Accardo} \emph{et~al.}, ``Flight test of a radar-based tracking system for
  uas sense and avoid,'' \emph{IEEE Trans. Aerosp. Electron. Syst.}, vol.~49,
  no.~2, pp. 1139--1160, 2013.

\bibitem{4357952}
J.~{Gu} and P.~{Wei}, ``Joint svd of two cross-correlation matrices to achieve
  automatic pairing in 2-d angle estimation problems,'' \emph{IEEE Antennas
  Wireless Propag. Lett.}, vol.~6, pp. 553--556, 2007.

\bibitem{4435061}
Y.~{Wu} and H.~C. {So}, ``Simple and accurate two-dimensional angle estimation
  for a single source with uniform circular array,'' \emph{IEEE Antennas
  Wireless Propag. Lett.}, vol.~7, pp. 78--80, 2008.

\bibitem{6547987}
A.~{Clemente} \emph{et~al.}, ``Wideband 400-element electronically
  reconfigurable transmitarray in x band,'' \emph{IEEE Trans. Antennas
  Propag.}, vol.~61, no.~10, pp. 5017--5027, 2013.

\bibitem{6898043}
B.~{Ku} \emph{et~al.}, ``A 77–81-ghz 16-element phased-array receiver with
  $\pm {\hbox{50}}^{\circ}$ beam scanning for advanced automotive radars,''
  \emph{IEEE Trans. Microw. Theory Tech.}, vol.~62, no.~11, pp. 2823--2832,
  2014.

\bibitem{7544459}
B.~{Avser}, J.~{Pierro}, and G.~M. {Rebeiz}, ``Random feeding networks for
  reducing the number of phase shifters in limited-scan arrays,'' \emph{IEEE
  Trans. Antennas Propag.}, vol.~64, no.~11, pp. 4648--4658, 2016.

\bibitem{1143830}
R.~Schmidt, ``Multiple emitter location and signal parameter estimation,''
  \emph{IEEE Trans. Antennas Propag.}, vol.~34, no.~3, pp. 276--280, Mar 1986.

\bibitem{32276}
R.~Roy and T.~Kailath, ``Esprit-estimation of signal parameters via rotational
  invariance techniques,'' \emph{IEEE Trans. Acoust., Speech, Signal Process.},
  vol.~37, no.~7, pp. 984--995, Jul 1989.

\bibitem{4350230}
J.~{Li} and P.~{Stoica}, ``Mimo radar with colocated antennas,'' \emph{IEEE
  Signal Process. Mag.}, vol.~24, no.~5, pp. 106--114, 2007.

\bibitem{4358016}
J.~{Li} \emph{et~al.}, ``On parameter identifiability of mimo radar,''
  \emph{IEEE Signal Process. Lett.}, vol.~14, no.~12, pp. 968--971, 2007.

\bibitem{5456168}
P.~{Pal} and P.~P. {Vaidyanathan}, ``Nested arrays: A novel approach to array
  processing with enhanced degrees of freedom,'' \emph{IEEE Trans. Signal
  Process.}, vol.~58, no.~8, pp. 4167--4181, 2010.

\bibitem{5609222}
P.~P. {Vaidyanathan} and P.~{Pal}, ``Sparse sensing with co-prime samplers and
  arrays,'' \emph{IEEE Trans. Signal Process.}, vol.~59, no.~2, pp. 573--586,
  2011.

\bibitem{7277022}
J.~{Lee}, J.~{Lee}, and J.~{Woo}, ``Method for obtaining three- and
  four-element array spacing for interferometer direction-finding system,''
  \emph{IEEE Antennas Wireless Propag. Lett.}, vol.~15, pp. 897--900, 2016.

\bibitem{intAOA}
P.~Q.~C. Ly, \emph{Fast and unambiguous direction finding for digital radar
  intercept receivers}.\hskip 1em plus 0.5em minus 0.4em\relax Thesis. The
  University of Adelaide, School of Electrical and Electronic Engineering,
  2013.

\bibitem{8789717}
W.~{Stevers} \emph{et~al.}, ``Direction-of-arrival estimation using a low-cost,
  portable, software-defined-radio-based phase interferometry system [education
  corner],'' \emph{IEEE Antennas Propag. Mag.}, vol.~61, no.~4, pp. 78--84,
  2019.

\bibitem{aps2020}
S.~{Vakalis} and J.~A. {Nanzer}, ``A space-time modulated distributed antenna
  array for multiple target angle estimation,'' in \emph{2020 IEEE
  International Symposium on Antennas and Propagation and USNC-URSI Radio
  Science Meeting}, 2020.

\bibitem{8493305}
S.~{Vakalis}, E.~{Klinefelter}, and J.~A. {Nanzer}, ``Angle estimation using
  wideband frequency modulation and an active distributed array,'' \emph{IEEE
  Microw. Wireless Compon. Lett.}, vol.~28, no.~11, pp. 1059--1061, 2018.

\bibitem{7836351}
J.~A. {Byford} \emph{et~al.}, ``Demonstration of rf and microwave passive
  circuits through 3-d printing and selective metalization,'' \emph{IEEE Trans.
  Compon. Packag. Technol.}, vol.~7, no.~3, pp. 463--471, March 2017.

\end{thebibliography}

\end{document}